# A microfluidic chip and its use in characterising the particle-scale behaviour of Microbial-Induced Calcium Carbonate Precipitation (MICP)

Yuze Wang[1*], Kenichi Soga[2], Jason T. DeJong[3], Alexandre J. Kabla[1]

[1]Department of Engineering, University of Cambridge, Cambridge, CB2 1PZ, United Kingdom
[2]Department of Civil and Environmental Engineering, University of California, Berkeley, CA 94720, United States
[3]Department of Civil and Environmental Engineering, University of California, Davis, CA 95616, United States
Correspondence: yw369@cam.ac.uk; ORCID: 0000-0003-3085-5299.

**Abstract**

Microbial-Induced Calcium Carbonate Precipitation (MICP) is an innovative ground improvement technique which can enhance the strength and stiffness of soils and can also control their hydraulic conductivity. These engineering properties of MICP-treated soils are affected by the particle-scale behaviour of the precipitated calcium carbonate, i.e. composition, amount and distribution, which are controlled by the MICP process occurring at the particle-scale. In this study, we designed and fabricated a microfluidic chip to improve our understanding of MICP at particle-scale by observing the behaviour of bacteria and calcium carbonate crystals during this process. We found that bacteria became evenly distributed throughout the microfluidic chip after the injection of bacterial suspension, grew during bacterial settling, and detached during the injection of cementation solution. Bacteria aggregated during the cementation solution injection, and calcium carbonate crystals formed at narrow pore throats or open pore bodies either during or after cementation solution injections.

## 1. Introduction

Microbial-Induced Calcium Carbonate ($CaCO_3$) Precipitation (MICP) is a promising soil stabilisation technique which has been extensively studied for applications such as ground improvement, liquefaction prevention, dam safety, erosion prevention and slope stabilisation

(van Paassen, 2009; DeJong et al., 2013; Martinez et al., 2013; Montoya et al., 2013; Jiang et al., 2017). Ureolysis-based MICP is the most widely studied MICP process and involves the hydrolysis of urea by the urease produced by active microorganisms (eq. 1), resulting in the formation of $CaCO_3$ (eq. 2).

$$CO(NH_2)_2 + 2H_2O \xrightarrow{Urease} 2NH_4^+ + CO_3^{2-} \qquad 1$$

$$Ca^{2+} + CO_3^{2-} \rightarrow CaCO_3(s) \qquad 2$$

The precipitated $CaCO_3$ can bond soil particles and is capable of enhancing the strength and stiffness of soils. In addition, the precipitated $CaCO_3$ does not fully fill the pores, thus enabling MICP to be used to modify the hydraulic conductivity of soils by controlling the amount of carbonate precipitation (DeJong et al., 2006; DeJong et al., 2010; Al Qabany and Soga, 2013). The engineering properties of MICP-cemented sandy soils, such as strength, stiffness and permeability, are affected by $CaCO_3$ content (Whiffin et al., 2007; van Paassen et al., 2010; Martinez et al., 2013; Zhao et al., 2014). Although correlations between the $CaCO_3$ content and soil engineering properties have been discussed among different studies, the engineering properties of cemented sandy soils containing the same amount of $CaCO_3$ vary significantly (Wang et al., 2017). Differences in the properties of $CaCO_3$ crystals at particle scale could be a reason for these variations.

The particle scale properties of $CaCO_3$ crystals in MICP-treated soils, such as shape, size, amount and distribution, have been extensively studied by scanning electron microscopy (SEM). The shapes of $CaCO_3$ crystals were found to be either spherical or prismatic, which might represent vaterite or calcite, respectively (van Paassen, 2009; Al Qabany et al., 2012; Zhao et al., 2014). The size and number of $CaCO_3$ crystals were different in samples treated with different chemical concentrations and injection rates (Al Qabany et al., 2012). In terms of distribution, precipitates either coated the surface of sand particles or formed bridges between the sand particles as agglomerated spherical crystals (DeJong et al., 2010). Traditional particle scale observation methods such as SEM are capable of capturing the crystals after MICP treatment but not during the MICP process itself. In addition, although environmental scanning electron microscopy (ESEM) has previously been used to observe bacteria in wet samples (Little et al., 1991), no research so far has reported the capability of SEM or ESEM to monitor the behaviour of live bacteria under depressurised conditions. Therefore, a new technology that can observe the MICP process and the behaviour of both bacteria and carbonate crystals at particle scale under saturated conditions remains to be developed.

In this study, a microfluidic device containing a porous medium was designed based on a cross-sectional image of solidified and sectioned 3D Ottawa sandy soil specimen and fabricated using polydimethylsiloxane (PDMS). The inner surface of the porous medium representing soil particles was treated with plasma to make it hydrophilic and negatively charged to simulate the surface properties of sandy soil particles. Using an injection system, the flow rate was controlled to be the same level as the flow rate used during lab-scale MICP treatment. Experiments were conducted to observe the particle scale behaviour of bacteria and $CaCO_3$ crystals during MICP treatment, starting with the injection of bacterial suspension and followed by multiple injections of cementation solution. Images were taken using a computer-controlled microscope both during and between these injections.

**2. Background**

Microfluidics enables the manipulation of small amounts of fluid using channels with dimensions of tens to hundreds of micrometres (Whitesides, 2006). The channels are built into a small device which is normally called a microfluidic chip. Initially stemming from microanalytical methods and microelectronic circuits in the early 1990s (Whitesides, 2006), the field has expanded dramatically in the past decade. This is largely due to the introduction of soft lithography, an easily accessible fabrication technique which is used to create microfluidic chips based on patterned elastomeric polymers. Polydimethylsiloxane (PDMS) is the most commonly used material to date owing to its low cost, easy fabrication, flexibility, and optical transparency (Sia and Whitesides, 2003; Weibel et al., 2007). In addition, their surface properties such as wetting properties (hydrophobic and hydrophilic) and charges (positive and negative) can be varied using different surface treatment technologies (Bodas and Khan-Malek, 2007; Wong and Ho, 2009; Zhou et al., 2010, 2012). Microfluidics has revolutionised fundamental and applied research to investigate and observe small-scale physical, chemical, and biological processes in the fields of chemical, biological, medical and environmental engineering (Tohidi et al., 2001; Sia and Whitesides, 2003; Zhang et al., 2010; Long et al., 2013; Jaho et al., 2016). The introduction of microfluidic platforms into the study of particle-scale behaviour of MICP for applications in Geotechnical Engineering is very recent (Wang et al., 2017) and in many cases still in its infancy. In this study, we designed and fabricated a microfluidic chip that replicates key features of the porous matrix of sandy soil by using a collection of irregular-shaped pillars to mimic sand grains and the pores between them. In addition, we used microfluidics to characterise the particle-scale MICP process.

## 3. Materials and Methods

### 3.1 Design of a Mask for Fabricating the Microfluidic Chips

The designed microfluidic chip contains an inlet, upstream flow distribution channels, a porous medium, downstream flow distribution channels, and an outlet. The two-dimensional design of the microfluidic chip is shown in Fig. 1(a). The inlet and outlet are used to inject and to let out the bacterial suspension and cementation solution, respectively. The flow distribution channels distribute the flow evenly into and out of the porous medium. The porous medium in the middle of the microfluidic chip represents the porous soil matrix. It is 15 mm in length and width, and 50 µm deep. It contains a matrix of irregularly-shaped pillars with pores to enable flow between the pillars. The shape, size and distribution of the pillars, which represent sandy soil particles, were designed based on a cross-sectional image of a solidified and sectioned 3D Ottawa 30-50 sandy soil specimen (Fig. 1(b) in Yang, 2005). In real soil, fluid flows through pores between the soil particles. However, in the cross-sectional image of the soil (Fig. 1(b) and (c)), some of the particles are connected to each other, meaning that when this image is used to make a microfluidic chip with irregular pillars, fluid cannot flow between these pillars. Therefore, contacts between the particles (Fig. 1(c)) were manually modified to create open channels (Fig. 1(d)). Most of the distances between two adjacent pillars were shorter than 50 microns, roughly equivalent to the depth of the microfluidic chip. The calculated porosity based on the geometric design was approximately 0.40. The designed mask was printed by Micro Lithography Services (Chelmsford, United Kingdom) on a transparent film for the fabrication of the microfluidic chips.

**3.2 Fabrication of the PDMS Microfluidic Device**

In this study, standard photolithography techniques were used to fabricate the master for the microfluidic device, mainly using a silicon wafer (MicroChemicals GmbH, Germany), SU-8 3025 (MICRO CHEM, US) and EC solvent (Dow Chemical Company, Denmark). The silicon wafer was coated with SU-8 3025, which is a photoresist (PR) polymer. The coating was performed using a spin coater, the speed of which controls the thickness of SU-8 (50 µm in this study). The SU-8-coated silicon wafer was heated to be solid after coating and was then selectively exposed to UV light by placing the mask directly above the wafer. The areas of PR which were not exposed to UV light were weakened and removed using EC solvent. The fabricated silicon master is shown in Fig. 2.

The fabrication of the microfluidic device was conducted after the fabrication of the silicon master. A 10:1 w/w mixture of PDMS base and curing agent (Sylgard 184, Ellsworth Adhesives) was poured over the master and cured in an oven at 65 °C for 5 hours, after which the cured PDMS stamp was peeled from the silicon master. After the inlet and the outlet holes were cut with a Harris Uni-Core 0.75 mm punch (Ted Pella, Inc.), the PDMS chips were irreversibly sealed with clean PDMS bottom slips by exposing them to air plasma (Harrick Plasma). PDMS consists of repeated units of $-O-Si(CH_3)_2-$, with methyl groups ($-CH_3$) being converted to silanol groups ($Si-O-H$) after exposure to oxygen plasma (Zhou et al., 2010).

**3.3 MICP Testing Process**

3.3.1 Preparation of Bacterial Suspension and Cementation Solution

*Sporosarcina pasteurii* [American Type Culture Collection (ATCC) 11859] was used in all experiments due to its ability to synthesise urea (Al Qabany et al., 2012). This bacterial suspension was cultivated in Ammonium-Yeast Extract ($NH_4$-YE) liquid medium containing 0.13 mol $L^{-1}$ tris buffer, 10 g $L^{-1}$ ammonium sulphate [$(NH_4)_2SO_4$], and 20 g $L^{-1}$ yeast extract (Soon et al., 2014). *S. pasteurii* cells were grown at 30°C for 5.5 - 16 hours to obtain an optical density at 600 nm ($OD_{600}$) of 0.8 - 4.0. Optical density, measured using a spectrophotometric method (Widdel, 2007), indicates the turbidity of the bacterial suspension, which correlates with both bacterial concentration and bacterial size (Zapata and Ramirez-Arcos, 2015). The cementation solution for MICP treatment contained 0.25 M calcium chloride dihydrate (CaCl$_2$·2H$_2$O), 0.375 M urea [CO(NH$_2$)$_2$], 0.187 M ammonium chloride (NH$_4$Cl), 0.0252 M sodium bicarbonate (NaHCO$_3$), and 3 g/L nutrient broth. All chemicals used were of Analytical Reagent grade.

3.3.2 Particle-scale Experiments using Microfluidic Chips

Fig. 3 shows the schematic of the microfluidic chip experiments. PDMS microfluidic chips were fully saturated with deionised (DI) water prior to the microfluidic experiments. Bacterial suspension was injected into the microfluidic chips using a syringe pump at controlled flow rates and durations. Bacteria were then given time to settle and attach to the PDMS surfaces within the chip. Subsequently, the syringe containing the bacterial suspension was displaced by a syringe containing cementation solution and, in the meantime, a new tubing was also replaced. The process of injecting the cementation solution was the same as the process of injecting bacteria, with different flow rates, volumes and numbers of injections being trialled. Images of the samples were taken at different time points.

**3.4 Characterisation of the particle-scale MICP process**

Experiments were conducted to characterise bacterial distribution after injection (Test 1), bacterial growth during bacterial settling (Test 2), bacterial detachment during the injection of cementation solution (Tests 3 and 4), and crystal growth during and after the injection of cementation solution (Tests 5 and 6). The bacterial, chemical and injection parameters of the tests are summarised in Table 1 and Table 2. The OD$_{600}$ of the bacterial suspension was measured immediately before the bacterial injection. The amount of fluid perfused through the microfluidic chip was measured as the volume injected divided by the pore volume of the device (PV) (Tables 1 and 2). When observing bacterial transport during the injection of bacterial suspension, bacterial suspension was injected at a flow rate of 5.6 PV/h to allow enough time for images of the sample to be taken at different locations and at different time points during the injection. The settling time is the time interval between the completion of bacterial suspension injection and the start of the injection(s) of cementation solution (Table 1). In Tests 4 and 6, multiple injections of cementation solution were applied, and the time intervals between every two injections are shown in Table 2.

### 3.5 Data Acquisition and Image Analysis

All images were acquired with an Axio Observer Z1 research microscope. The microscope is equipped with an automated stage (Prior Scientific Instrument), a black and white camera (Hamamatsu C11440-22CU) and an optical light power source connected to a computer and controlled by Zeiss AxioVision image analysis software. For tests that investigate bacterial numbers, *S. pasteurii* in the micromodel were imaged using phase field illumination and 10× inverted objectives (with image resolutions of 0.65 µm/pixel). When studying the average number of bacteria in the microfluidic chip, the number of bacteria in the pores near the inlet, in the middle and near the outlet of the porous medium were counted and averaged to determine the average number of bacteria present throughout the porous medium. When investigating the change in the number of bacteria present at different positions, the number of bacteria in two pores closer to the inlet, two pores at the middle and two pores closer to the outlet were counted, after which the average number of bacteria at each of these positions were counted to represent the number of bacteria present at the left-hand side, middle and the right-hand side of the microfluidic chip. For tests that investigate $CaCO_3$ size and amount, images were taken using bright field illumination and 10× inverted objectives (with image resolutions of 0.65 µm/pixel). Images were analysed using Zeiss Axio Vision image analysis software to determine the areas of the corresponding crystals.

### 4. Results and Discussion

#### 4.1 Observations during Bacterial Injections

A uniform distribution of bacteria is thought to be the most important requirement for achieving uniform calcium carbonate precipitation (Martinez et al., 2013). During bacterial injection, bacteria transport through the porous medium by advection and dispersion. Due to the inhomogeneous distribution of the pores and particles in the porous medium, some of the bacteria are transported through the chip faster than others. Even though dispersion can help to achieve a more homogenous distribution of bacteria in the porous medium, the injection volume should be higher than 1 PV to achieve this. To check that flow is smooth and to check whether a uniform bacterial distribution can be obtained after the injection, in Test 1, after the microfluidic chip was fully saturated with DI water, the change in the number of bacteria with time and space in the porous medium during the injection of bacteria was investigated. The direction of the injection of bacterial suspension into the microfluidic chip was from left to right in the photos shown in this paper. Three areas were imaged - at the left-hand side, middle and the right-hand side of the porous medium of the microfluidic chip (Fig. 4(a)). Time-series images of these three areas were taken during the 1.5 PV injection of bacterial suspension and the numbers of bacteria in two squares in each of these images were counted. Magnified images of squares A, C and E in Fig. 4(a) and the quantification of bacteria in the six squares at the completion of injecting 0.25, 0.5, 0.75, 1.0, 1.25 PV and 1.5 PV during the bacterial injection are shown in Fig. 4(b) and (c), respectively. The bacterial density in the pore space was higher closer to the injection face and increased as the number of pore volumes injected at all measurement locations increased. After injecting 1 PV of bacterial suspension, the number of bacteria on the left side of the porous medium was higher than in the middle, which in turn was higher than on the right-hand side of the porous medium closest to the outlet. Differences in numbers of bacteria between these three areas became smaller as the PV number of injected bacteria increased. The minimum pore volume of bacterial suspension required to achieve a relatively homogeneous distribution of bacteria inside the porous medium was between 1.0 and 1.25 PV. Therefore, in most of the experiments in this study, 1.25 PV of bacterial suspension was injected into the microfluidic chips to obtain a uniform distribution of bacterial cells. However, as bacterial cells contain flagella that enable them to move through fluids, a homogenous distribution of bacterial cells may also be achieved by allowing sufficient time for the bacteria to distribute within the media due to their motility.

**4.2 Observations during Bacterial Settling**

Bacterial growth experiments have normally shown that under optimised conditions, bacterial growth follows six phases: lag phase, starting phase, exponential phase, slow-down phase, stationary phase and death phase (Widdel, 2007). The initial lag phase is followed by a rapid increase in the bacterial population, normally referred to as the exponential growth phase (Widdel, 2007; Garrett et al., 2008). In most previous studies, bacteria were cultivated to exponential phase (or early stationary phase) before being injected into the soil matrix, after which the bacterial number *in situ* was not assessed. In theory, bacteria exhibiting exponential growth still have the ability to multiply (Widdel, 2007; Garrett et al., 2008). Test 2 was therefore conducted to observe bacterial growth during bacterial settling. In Test 1, the $OD_{600}$ for *S. pasteurii* suspension reached a value of 4.0. Because the $OD_{600}$ of the bacterial suspension in Test 2 was 0.8 prior to the injection of bacterial suspension and the growth medium was the same as that used in Test 1, it was assumed that the bacterial suspension contained sufficient nutrients for bacteria to grow further. Optical microscope images of bacteria in three pores of the microfluidic chip at $t_0$, $t_0 + 2$ h and $t_0 + 4$ h during the settling of bacteria are shown in Fig. 5(a). $t_0$ was about 10 minutes after the completion of the bacterial injection. The number of bacteria in each image depicted in Fig. 5(a) was counted, with the results shown in Fig. 5(b). Bacteria with an $OD_{600}$ of 0.8 grew during bacterial settling, with bacterial numbers increasing by 80% every two hours (Fig. 5(b)). Based on the parameters given by Leclerc et al. (2003) and the thickness of the PDMS in this study (6 mm), the estimated rate of oxygen diffusion into the porous medium of the chip is $4.8 \times 10^{-5}$ mol per day. This oxygen level is higher than the amount of oxygen required for aerobic bacteria such as *E. coli* to grow (Varma and Palsson, 1994). *S. pasteurii* is also an aerobic bacterial strain, and is unable to grow anaerobically (Martin et al., 2012). In Test 2, it is shown that *S. pasteurii* increased in number after being injected, which suggests that sufficient oxygen is available for them to grow. In addition, it is important to note that since bacterial growth is affected by several factors such as temperature, dissolved oxygen (DO), nutrient and pH (Garrett et al., 2008), the observations in this study may not fully represent bacterial growth in real soils. However, the direct observation of bacteria during the MICP process described in this study can be used in future studies to investigate the effect of environmental factors on bacterial growth and the effects of bacterial amount on the properties of calcium carbonate in a way that has not previously been possible.

### 4.3 Observations during the Injection of Cementation Solution

A significant fraction of the bacteria were flushed out during the injection of cementation solution. Optical microscope images of bacteria inside three pores on the left-hand side, middle and right-hand side of the microfluidic chip during the 1.25 PV injection of cementation solution are shown in Fig. 6(a). Bacteria in three representative images at the left-hand side, middle and the right-hand side of the microfluidic chip at injection PV numbers of 0, 0.25, 0.5, 0.75, 1.0 and 1.25 were counted, and the average results are shown in Fig. 6(b). The number of bacteria in the pores on the left side of the microfluidic chip continuously decreased throughout the time course of the injection, while the numbers of bacteria in the pores in the middle and right-hand side of the chip increased until around 0.25 to 0.5 PVs were injected, after which they began to decrease. This may be because during the injection of cementation solution, the bacterial cells at the left-hand side of the microfluidic chip were initially replaced by cementation solution which does not contain bacteria. Therefore, the bacterial density in the pores at the left-hand side decreased after the start of the injection. Meanwhile, during the injection of cementation solution, bacterial cells which became detached on the left-hand side of the chip were flushed through the chip and accumulated on the right-hand side. Therefore, the bacterial density in the right-hand side of the chip increased during the injection of the first 0.5 PV injection of cementation solution. After the cementation solution reached the middle of the chip, some of the bacteria in the middle of the chip became displaced by the cementation solution, resulting in a decrease in bacterial density at this location. Bacteria in the pores at the left-hand side, in the middle and at the right-hand side of the microfluidic chip stopped decreasing in number after around 0.75, 1 and 1.25 PV were injected, respectively. The number of bacteria which remained in the microfluidic chip after the injection of 1.25 PV of cementation solution was about half of the initial amount injected.

The bacterial population decreased with each subsequent injection of cementation solution, with the greatest reduction occurring during the first injection of cementation solution. Optical microscope images of bacteria in one of the pores at the centre of the microfluidic chip taken after bacteria were allowed to settle for 4.5 hours (before the first cementation solution injection), and after the first, third and twelfth injection of cementation solution during twelve sequential injections are shown in Fig. 7(a). Bacteria were counted after the bacteria were allowed to settle for 4.5 hours, and after each of the twelve cementation solution injections (Fig. 7(b)). About half of the bacteria were flushed out after the first 1.25 PV of cementation solution was injected. The number of bacteria continued decreasing until around the seventh cementation solution injection, after which the number of bacteria remained constant. The number of bacteria at the end of the series of cementation solution injections was about 20% of the number of bacteria present immediately after settling. Previous research indicates that bacteria tend to secrete adhesive structures such as flagella, pili, exopolysaccharides, and other matrix components to attach to the solid surface and live at the soild-liquid interface (Dunne, 2002). This process generates an adhesive force between bacterial and solid surfaces (Persat et al., 2015). After settling, a proportion of bacteria were still in suspension or loosely attached to an interface, but some bacteria successfully adhered. The first couple of injections essentially cleared the bacteria that were in suspension or failed to adhere.

**4.4 Formation of Cementation with Time**

Formation of cementation with time might occur during or after the injection of cementation solution. In Test 5, six pore volumes of cementation solution were continuously injected into one microfluidic chip at a flow rate of 5.6 PV/h, and images were taken from the start of the first until the sixth PV injection of cementation solution (Fig. 8). Bacterial aggregates were observed (Fig. 8 (a)), possibly because bacterial cell walls are negatively charged and could thus be bonded by the $Ca^{2+}$ cations of $CaCl_2$ present in the cementation solution (Rodriguez-Navarro et al., 2007 and El Mountassir et al, 2014). Crystals appeared either at the narrow pore throat, the bottom, or the side wall of the inner surface of the microfluidic chip (Fig. 8). After being formed, the nucleation of cementation continued growing during the injection process.

In Test 6, a staged process of injecting cementation solution was conducted. During this process, 20 sequential 1.25 PV of cementation solution were injected into the microfluidic chip after the completion of bacterial settling, at an interval of about 24 hours between injections. The injections of cementation solution were conducted over the course of about 20 days in total. An image showing cementation throughout the whole porous medium of the microfluidic chip after the completion of the final cementation solution injection is presented in Fig. 9(a). Representative images of crystals inside six pores of this porous medium at the completion of the final cementation solution injection are also shown in this figure. The shape of the crystals was mainly prismatic, suggesting that the crystals might be calcite (Al Qabany et al., 2012; Zhao et al., 2014). Images of pores A to F at the completion of the $4^{th}$, $8^{th}$, $12^{th}$, $16^{th}$ and $20^{th}$ injection of cementation solution are shown in Fig. 9(b). Pores A-C are narrow, whereas pores D-F are wide. Most but not all the crystals at the final stage of the injection process were formed at the completion of the 4$^{th}$ injection of cementation solution. Once the crystals were formed, they continued to grow during subsequent injections of cementation solution. The sizes of the five crystals in pore A at the completion of the 4$^{th}$, 8$^{th}$, 12$^{th}$, 16$^{th}$ and 20$^{th}$ injection of cementation solution are shown in Fig. 9(c). At the completion of the 4$^{th}$ cementation solution injection, only one crystal was formed, while the others were formed between the 4$^{th}$ and the 8$^{th}$ cementation solution injection. The sizes of the five crystals at the same time point were different, while they all continued growing during subsequent injections of cementation solution. The ratios of total area of the crystals in the pores with respect to the pore areas of the narrow and open pores after every 4$^{th}$ injection are shown in Fig. 9(d). The crystal area to pore area ratios in the narrow and open pores at the completion of the 4$^{th}$ cementation solution injection were about 0.06 and 0.05, respectively. The ratios increased gradually to about 0.45 and 0.38, respectively, until the completion of the final cementation solution injection. It should be noted that the ratio of crystal area to pore area in this study is not the same as the ratio of crystal volume to pore volume, as the crystals might not occupy the whole height of the microfluidic chip.

The cementation formed at narrow pore throats might represent the effective $CaCO_3$ required for increasing the strength and stiffness of soils (DeJong et al., 2010); all of the cementation inside soils decreases permeability, and large crystals at the open pore throat might correspond to the effective $CaCO_3$ required to reduce soil permeability (Al Qabany and Soga, 2013). The concentration of chemicals used in MICP treatment affects the size and distribution of crystals, which consequently affects the mechanical properties of MICP-treated samples (Al Qabany and Soga, 2013). However, to date, no research has been conducted to explain the reason behind this. The observation of crystal growth inside the porous medium may help to improve our understanding of the mechanisms responsible for the different spatial distribution, shape, sizes, and amount of $CaCO_3$ at particle scale.

**5. Conclusions**

In this study, transparent microfluidic chips were designed based on a cross-sectional image of Ottawa 30-50 sandy soil and were fabricated using PDMS. Particle-scale MICP tests were conducted using the microfluidic chips to investigate the behaviour of both bacteria and carbonate precipitates at different stages during the whole MICP process. Key observations are summarised as follows:

The porous medium of a soil matrix can be simplified and simulated using a microfluidic chip. The PDMS microfluidic chip is impermeable to water and the injection flow rate of both bacterial suspension and cementation solution in the chip can be kept at the same level as in macro-scale soil column experiments. Microfluidic chips made of PDMS are transparent, which enables bacteria and calcium carbonate crystals to be observed using an optical microscope. Unlike samples imaged using SEM, samples do not need to be under a depressurised condition when being imaged using an optical microscope. In addition, the use of an optical microscope enables bacteria and calcium carbonate crystals to be observed in a saturated condition. The image intensities of bacteria, water, PDMS chip, and crystals are different, which makes it very easy to distinguish them. The process of MICP inside the porous medium both during and after MICP treatment can be observed without breaking the samples.

Flow smoothly perfused through the porous medium in the microfluidic chip and bacteria were homogeneously distributed in the chip after 1.25 PV of bacterial suspension was injected. This observation supports the findings of Tobler et al. (2014) that the distribution of *S. pasteurii* in sandstone at a larger scale (1.8 - 7.5 cm) was fairly homogeneous after the initial bacterial injection when the bacterial $OD_{600}$ was either 0.5 or 1.0. However, as the transport of bacteria through a porous medium might be affected by factors such as bacterial density, the scale of the samples, the porosity of the porous medium, and the injection flow rate, we are currently conducting a further study to investigate the effects of these parameters on the transport of bacteria.

Exponential growth of bacteria inside the porous medium during bacterial settling and the detachment of bacteria during the cementation solution injection were observed in this study. The bacterial density reported in the literature was usually measured before bacteria were injected into soils (Al Qabany et al., 2012; Montoya et al., 2013). However, as the number of bacteria might change after the injection of bacterial suspension, the number of bacteria present during injections of cementation solution might affect the precipitation rate and morphology of carbonate crystals. Using the microfluidic chip, further studies can be conducted to investigate factors affecting bacterial growth properties and the relationship between the number of bacteria and the properties of carbonate crystals such as growth rate and morphology. In addition, as the distribution of bacteria might affect the distribution of carbonate precipitation, the microfluidic chip could also be used to investigate the correlation between the distributions of bacteria and carbonate crystals.

$CaCO_3$ crystal growth was observed during either one continuous injection or twenty sequential injections. The observation of $CaCO_3$ crystals during the MICP process has great advantages over traditional ways of studying MICP by which only the $CaCO_3$ content at the end of the MICP treatment process can be measured (van Paassen, 2009; Al Qabany et al., 2012; Zhao et al., 2014). Monitoring the evolution of $CaCO_3$ during the MICP process can be useful to (i) study the mechanisms responsible for the spatial distribution of $CaCO_3$ at particle scale; (ii) estimate possible reasons for the formation of different shapes, sizes, and amount of $CaCO_3$, and (iii) monitor $CaCO_3$ precipitation rate, which might be helpful for optimising MICP treatment parameters such as chemical concentration and injection frequencies.

**Acknowledgements**

The first author would like to acknowledge Cambridge Commonwealth, European and International Trust, and China Scholarship Council, who jointly funded this project. The authors would also like to thank Dr Thierry Savin and Dr Jerome Charmet for their support with soft lithography techniques, and Professor Tuomas Knowles for granting access to his laboratory in the Chemistry Department, University of Cambridge. Special thanks are also extended to Dr Osama Dawoud and Dr Ning-Jun Jiang for their advice and involvement in the discussion of results. The authors would also like to thank Fedir Kiskin for proofreading the manuscript.

**Figure captions**

Figure 1. (a) Microscope image of a printed mask for making the microfluidic chip; (b) cross-sectional image of a solidified and sectioned 3D Ottawa 30-50 sandy soil specimen by Yang (2005); (c) magnified image of the top-left square in image (b); (d) AutoCAD image of the realistic modification of image (c)

Figure 2. Photo of a fabricated silicon wafer master containing two modes

Figure 3. Schematic of the particle-scale experimental setup for studying MICP

Figure 4. Observations during the bacterial injection: (a) images taken left-hand side, in the middle and at the right-hand side of the porous medium to enable counting of bacteria; (b) time series images of squares A, C and E in (a) taken during the injection of 0.25 to 1.5 PV of bacterial suspension; (c) number of bacteria counted at different time points during the bacterial injection, with each data point representing the mean number of bacteria present in two counting areas (A and B for the left image, C and D for the middle image, E and F for the right image). Error bars correspond to standard errors

Figure 5. Observations during bacterial settling: (a) optical microscope images of bacteria in three pores of the microfluidic chip during bacterial settling at $t_0$, $t_0 + 2$ hours and $t_0 + 4$ hours after the bacterial injection; (b) bacteria counted during bacterial settling, with each data point representing the average value of the three counting areas and error bars corresponding to standard errors

Figure 6. Optical microscope images of bacteria inside three pores at the left (first row), middle (second row) and right (third row) of the microfluidic chip during injection of 1.25 PV of cementation solution; (b) bacterial numbers during the injection of 1.25 PV of cementation solution, with each data point representing the average value of the three counting areas and error bars corresponding to standard errors

Figure 7. (a) Optical microscope images of bacteria during the injection of cementation solution (CS). From left to right, images were taken after bacteria were allowed to settle for 4.5 hours (before the $1^{st}$ CS injection), after the $1^{st}$ CS injection, after the $3^{rd}$ CS injection, and after the $12^{th}$ CS injection; (b) bacterial numbers present after each of the twelve sequential CS injections, with each data point representing the average value of the three counting areas and error bars corresponding to standard errors

Figure 8. Time series microscopic images taken from the start of the first until the sixth PV injection of cementation solution: (a) bacterial aggregation and nucleation appear at a narrow pore throat; (b) and (c) nucleation at the inner surface of the microfluidic chip

Figure 9. (a) Image of the whole porous medium captured at the end of the staged injection procedure($20^{th}$ injection) and six magnified images of pores A-F; (b) images of pores A to F captured during the staged injection procedure at the completion of the $4^{th}$, $8^{th}$, $12^{th}$, $16^{th}$ and $20^{th}$ injections; (c) sizes of the five $CaCO_3$ crystals at position A at the completion of the $4^{th}$, $8^{th}$, $12^{th}$, $16^{th}$ and $20^{th}$ injections; (d) proportion of pore area occupied by crystals in the narrow and open pores at the completion of the $4^{th}$, $8^{th}$, $12^{th}$, $16^{th}$ and $20^{th}$ injections, with each data point

representing the average value of the three counting areas and error bars corresponding to standard errors

Table 1—Bacterial properties of the experiments

| Parameter | 1 | 2 | 3 | 4 | 5 | 6 |
|---|---|---|---|---|---|---|
| $OD_{600}$ | 4.0 | 0.8 | 1.6 | 1.6 | 0.8 | 1.6 |
| Injection PV | 1.5 | 1.5 | 1.25 | 1.25 | 1.5 | 1.25 |
| Flow rate, PV/h | 5.6 | 56 | 56 | 56 | 56 | 56 |
| Settling time, h | - | 4 | 3.5 | 4.5 | 4 | 3.5 |

Table 2—Chemical properties of the experiments

| Parameter | 1 | 2 | 3 | 4 | 5 | 6 |
|---|---|---|---|---|---|---|
| $Ca^{2+}$ concentration, M | - | - | 0.25 | 0.25 | 0.25 | 0.25 |
| Injection PV | - | - | 1.25 | 1.25 | 6 | 1.25 |
| Injection times | - | - | 1 | 12 | 1 | 20 |
| Flow rate, PV/h | - | - | 5.6 | 5.6 | 5.6 | 5.6 |
| Injection intervals, h | - | - | - | 6 | | 24 |

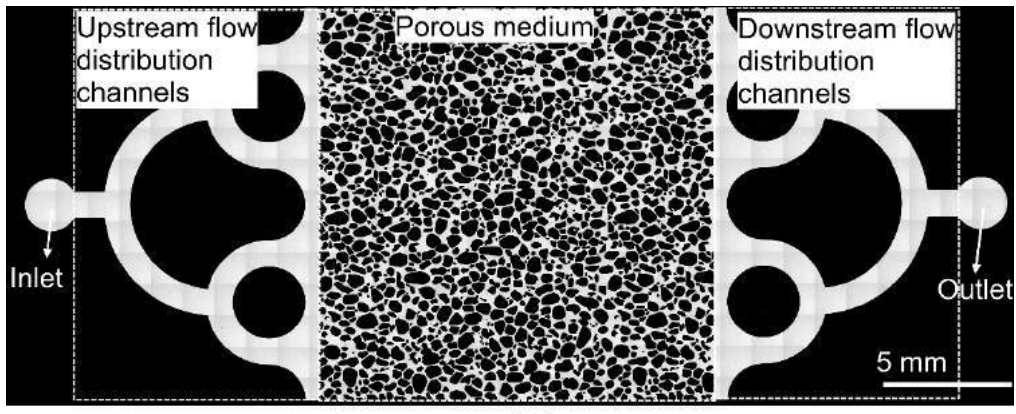

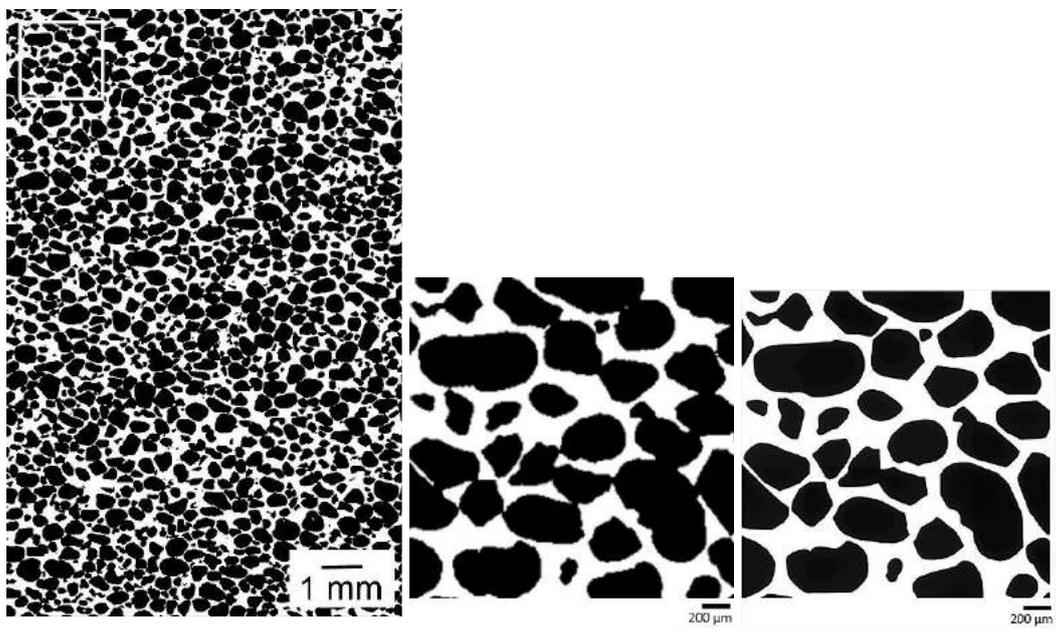

Figure 1

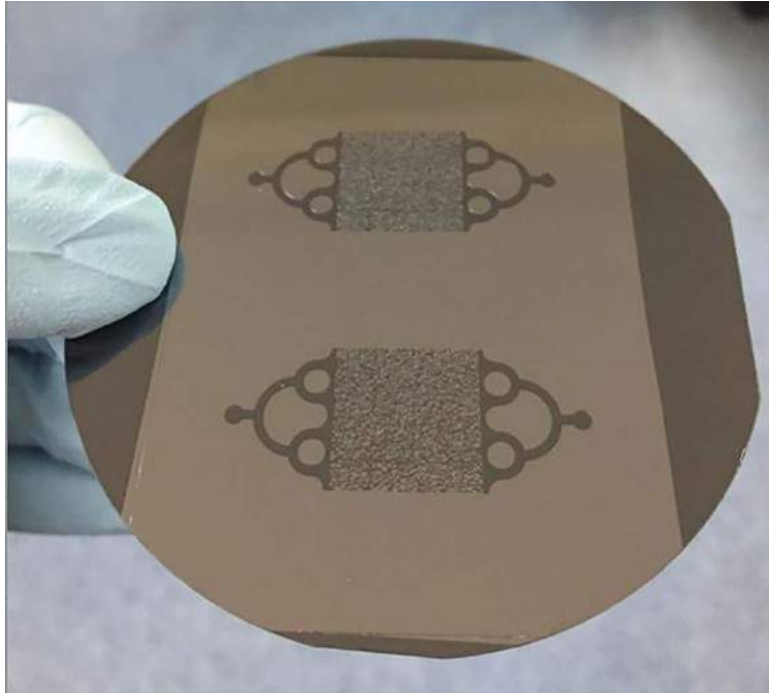

Figure 2

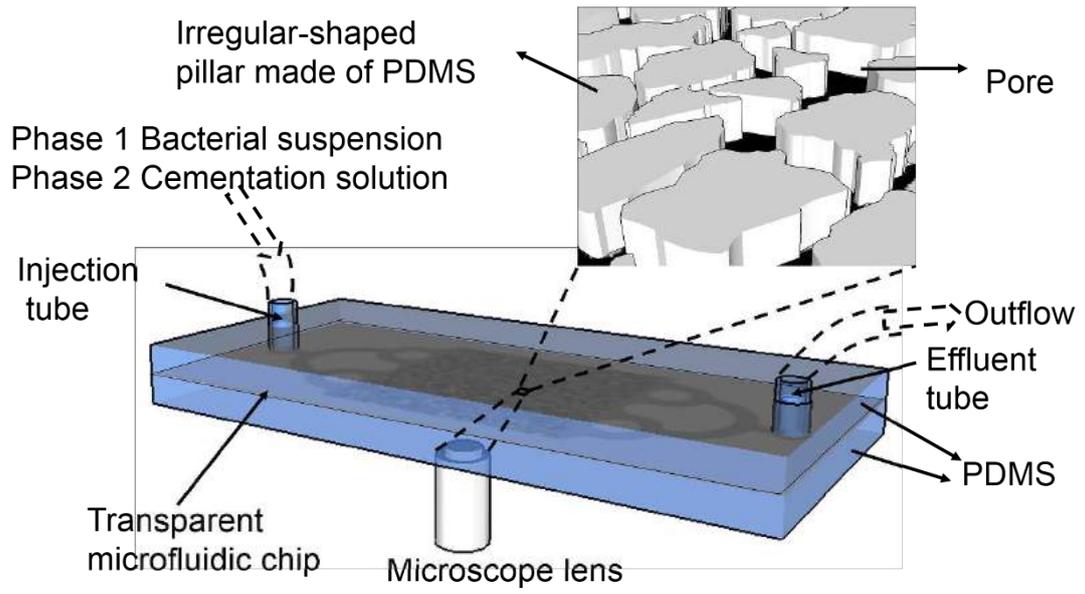

Figure 3

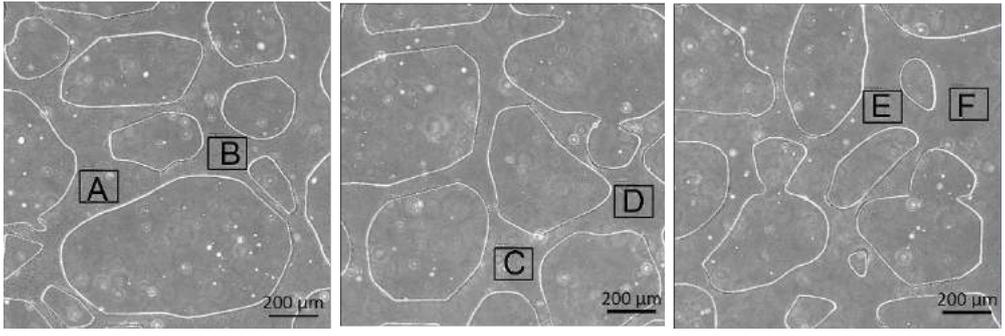

(a)

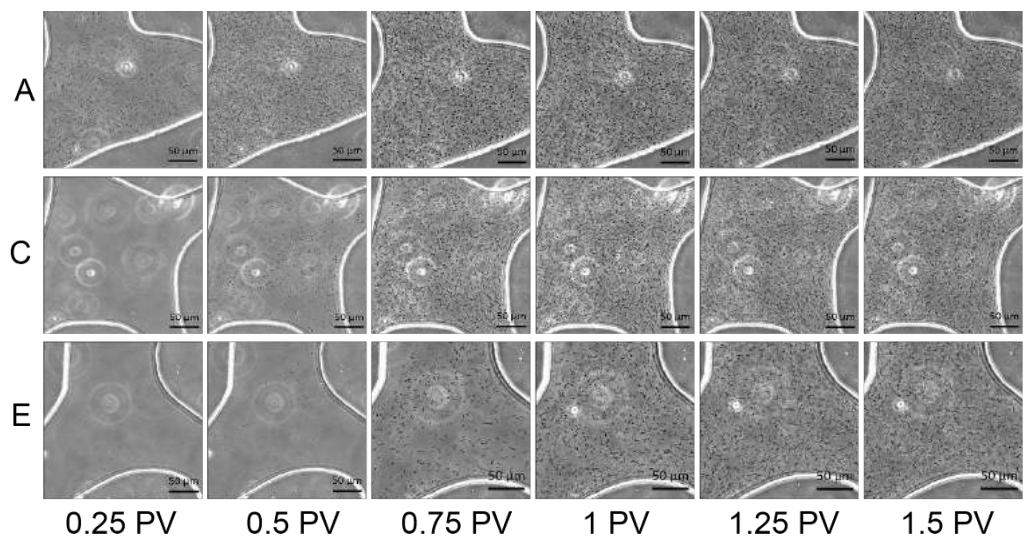

(b)

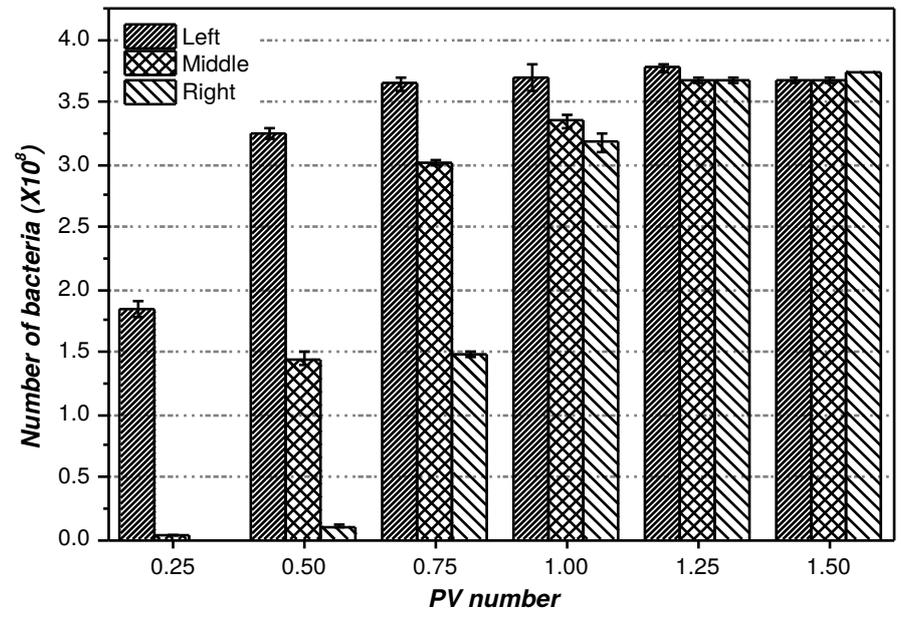

(c)

Figure 4

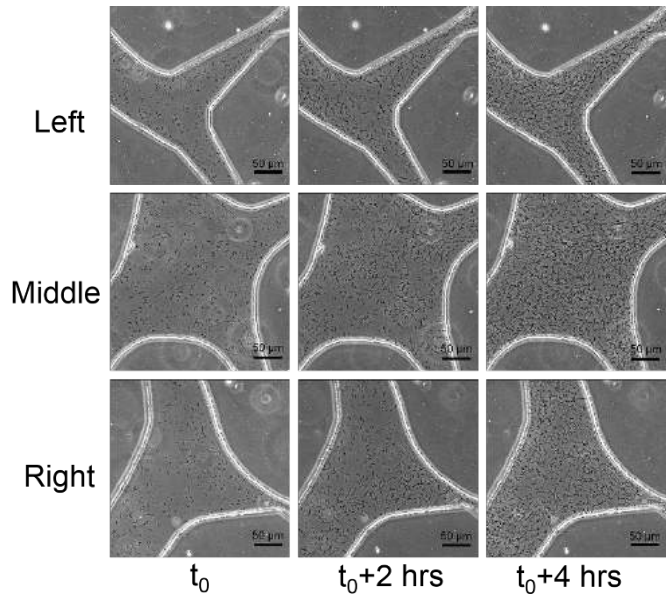

(a)

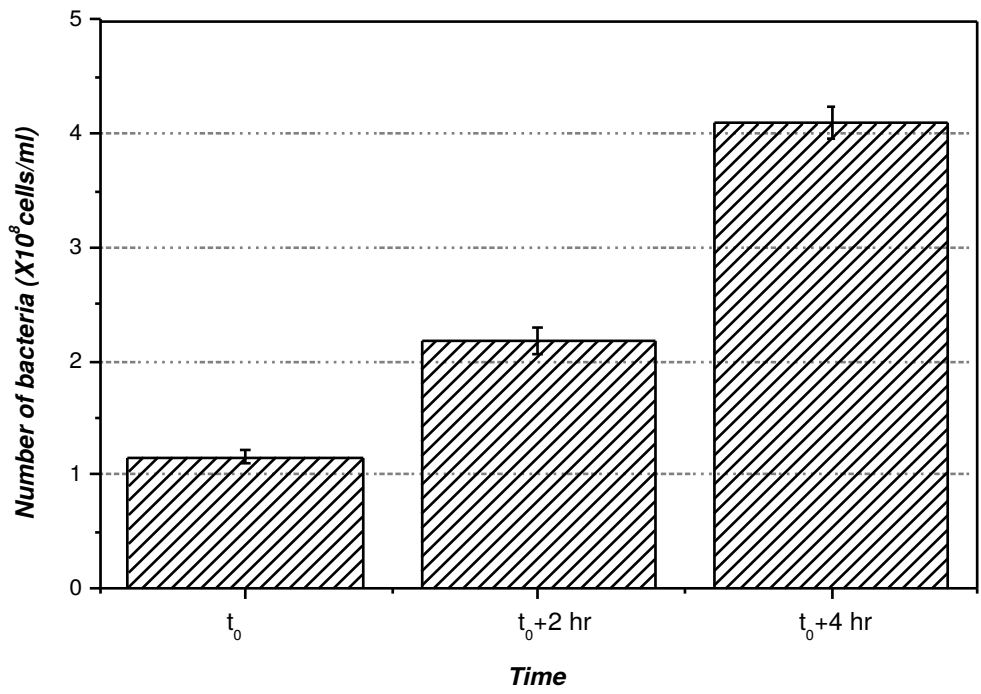

Figure 5

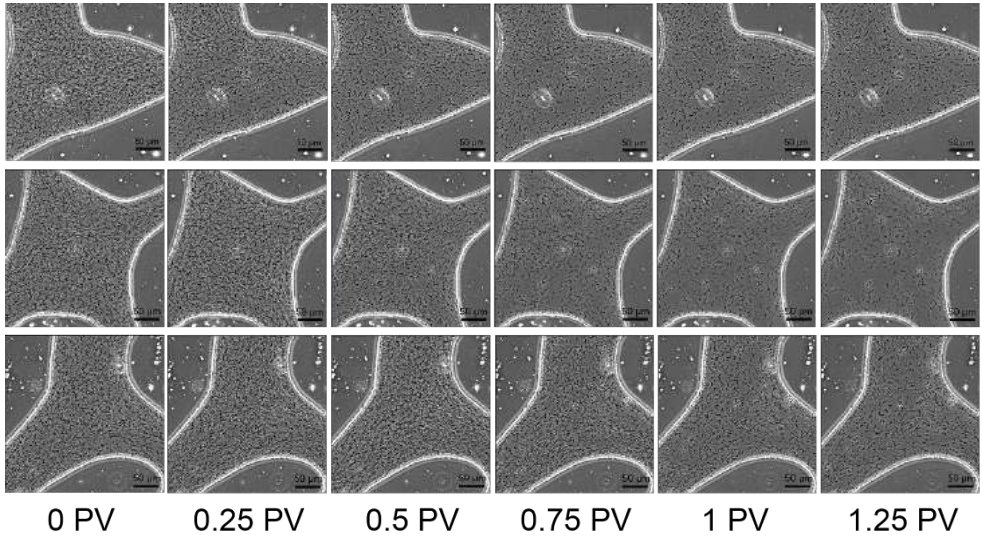

|  0 PV | 0.25 PV | 0.5 PV | 0.75 PV | 1 PV | 1.25 PV |

(a)

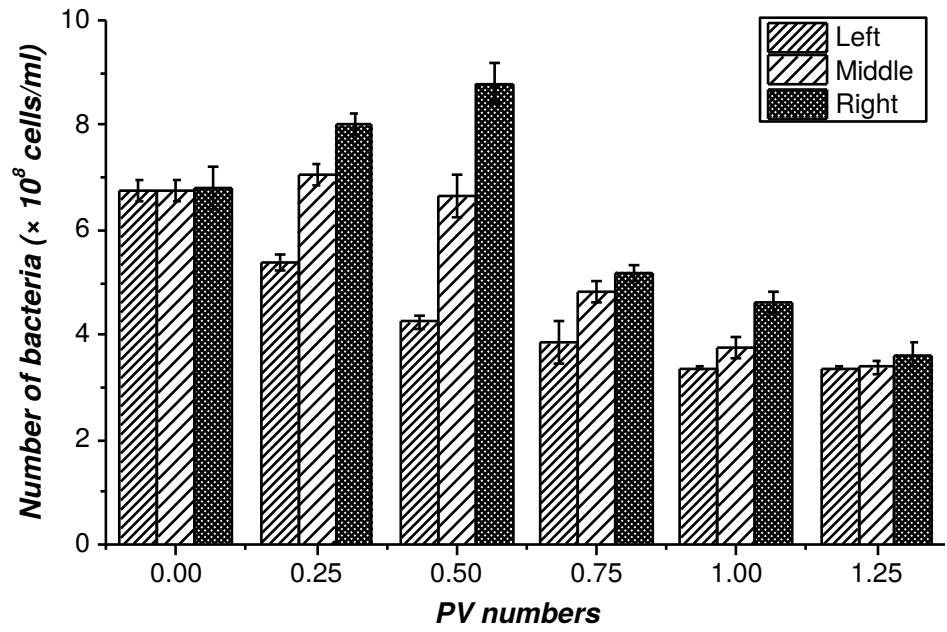

(b)

Figure 6

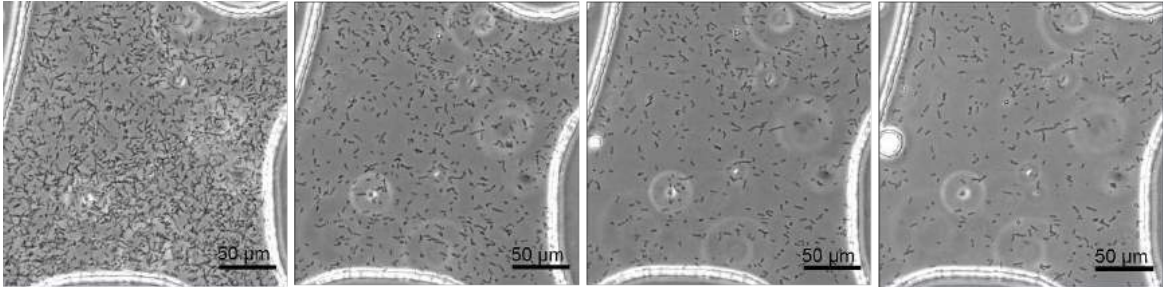

(a)

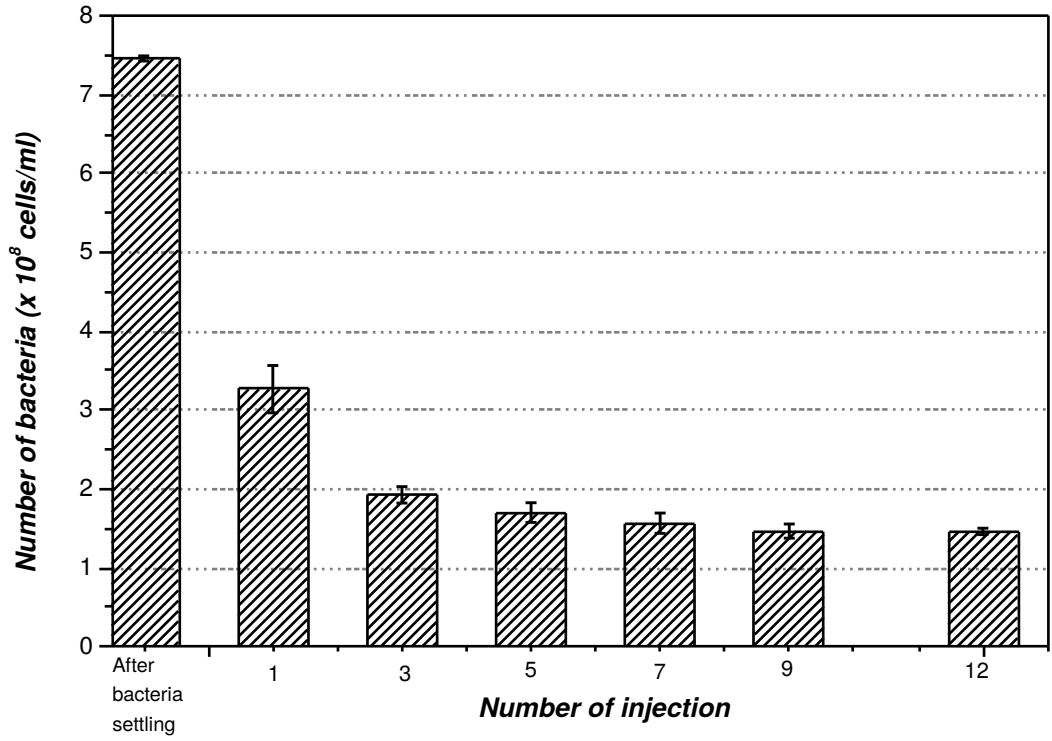

(b)

Figure 7

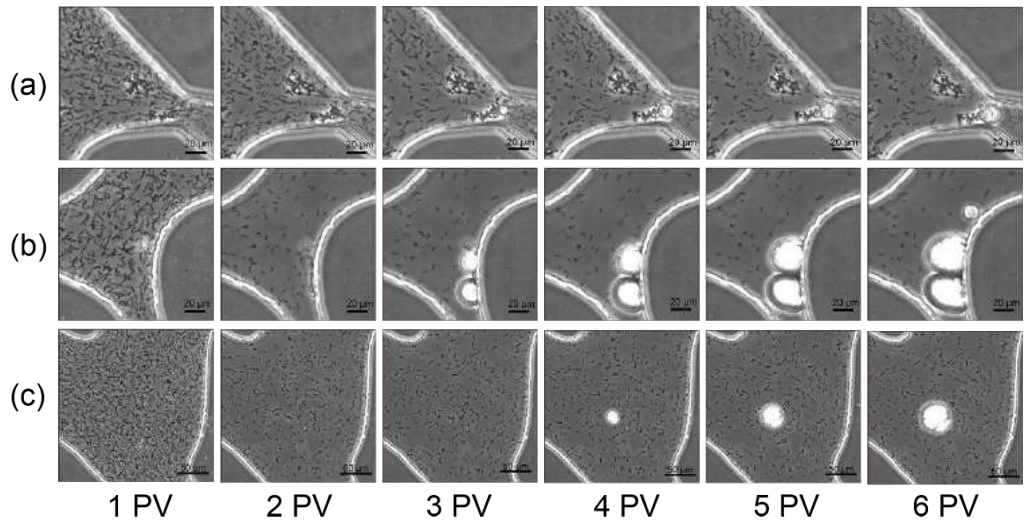

Figure 8

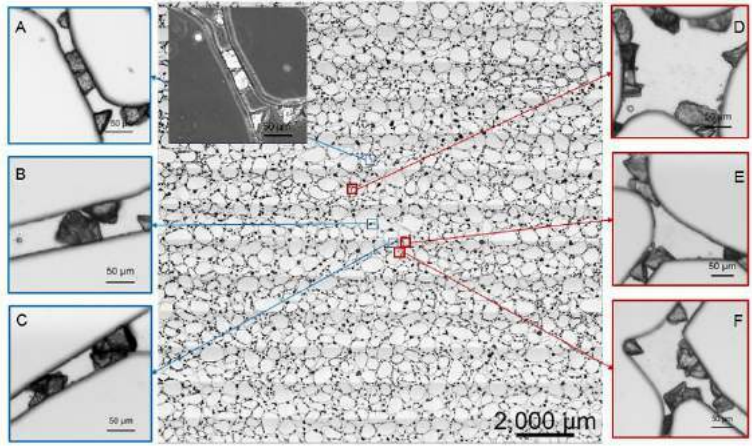

(a)

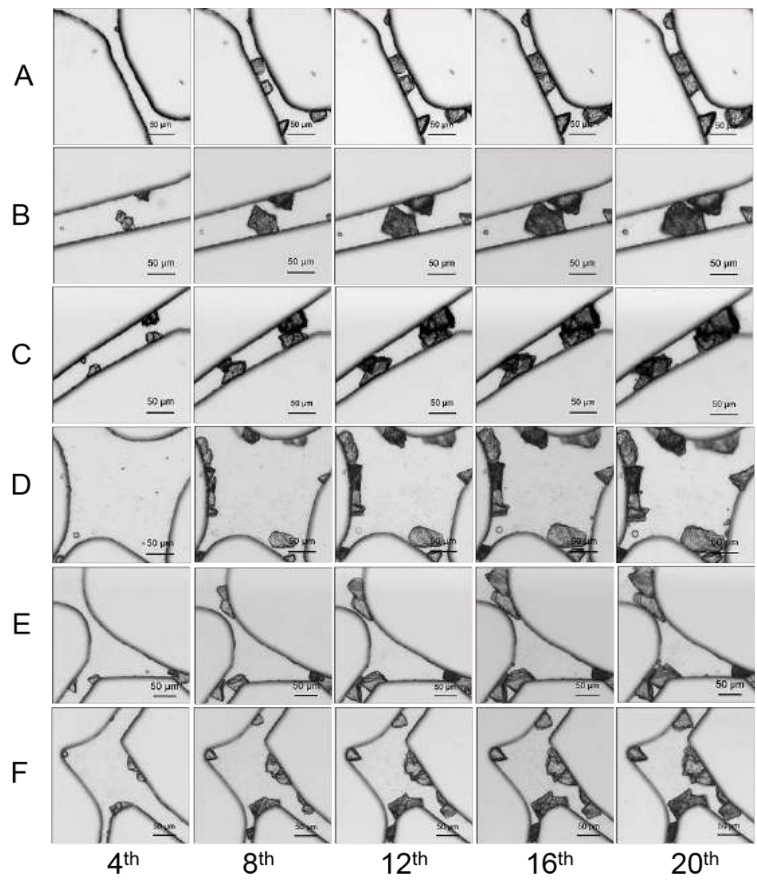

(b)

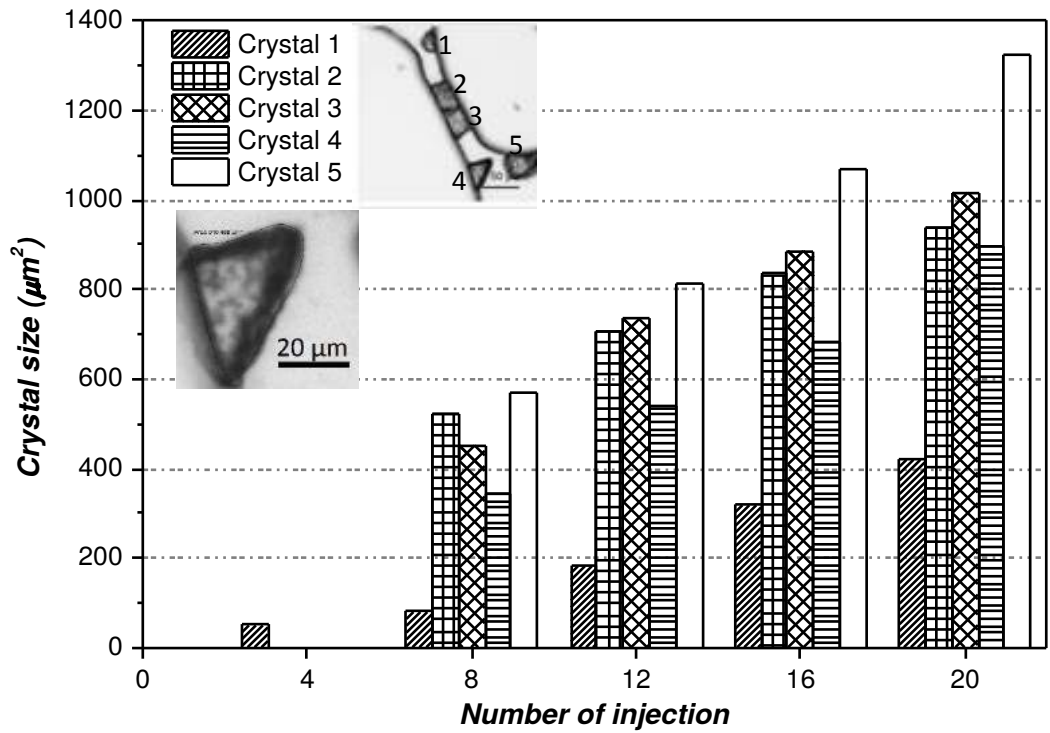

(c)

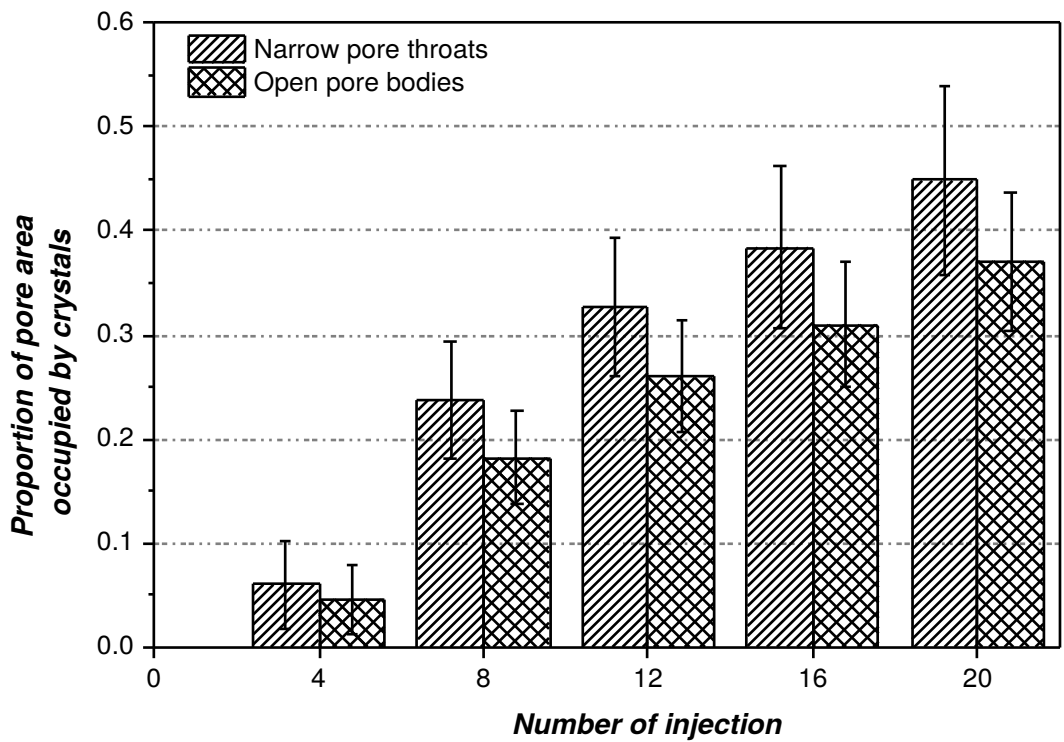

(d)

Figure 9